# Revised Elements and Blazhko Effect of the RR Lyrae Star AR Herculis

Erik Wischnewski[1]

[1]Bundesdeutsche Arbeitsgemeinschaft für Veränderliche Sterne e.V., Berlin, Germany

2016 June 8.

**Abstract**   AR Herculis is an RR Lyrae star of type RRab whose spectral type ranges from A7 to F3. Its visual brightness varies between 10.59 mag and 11.63 mag (V). By analysis of 34,000 measurements from the AAVSO International Database and additional data from the BAV, 116 times of maximum between 2003 and 2015 were derived. For the time span 2009 to 2015, a new period value of P = 0.46999746(±145) days was derived from the (O–C) analysis. Furthermore, the derived maximum timings indicate a strong Blazhko effect with a Blazhko period of $P_B$ = 32.062(±13) days. The evolution of the observed maximum brightness $V_{max}$, which decreased from 2009 to 2015 by 22 mmag/year, is also indicative of a Blazhko effect with $P_B$ = 29.528(±17) d. The present work investigates the processes in the frequently used (O–C) vs. $V_{max}$ diagram by running several simulations based on assumed and observational data.

## 1   Introduction

AR Herculis is an RR Lyrae star of type RRab. Its visual brightness varies between 10.59 mag and 11.63 mag (V) and its spectral type ranges from A7 to F3. The rise time to maximum (m–M) approximates 20% of the cycle length [Kholopov, 1988].

AR Herculis exhibits a strong Blazhko effect. Minimum and maximum brightness change considerably over the Blazhko cycle, which also holds true for the times of maximum brightness (O–C).

**Period |** For the period 1905 to 1954, Klepikova [1957] determined a period of P = 0.4700234 days ($E_0$ = HJD 2424794.274). For the time span 1954 to 1967, Lange [1969] derived a period of P = 0.469975 days ($E_0$ = HJD 2439692.683). The slightly longer period of P = 0.470028 days was given by Kholopov [1988] in the GCVS ($E_0$ = HJD 2441454.347). The elements from the GCVS (Equation 1) were adopted in the AAVSO International Variable Star Index, VSX. Feast et al. [2008] gave a period of P = 0.469981 days. Hübscher [2016a] determined the slightly longer period value of P = 0.46998756 days ($E_0$ = HJD 2454683.4366; Equation 2).

(1)   C = HJD 2441454.347   + 0.470028 ·E
(2)   C = HJD 2454683.4366 + 0.46998756 ·E

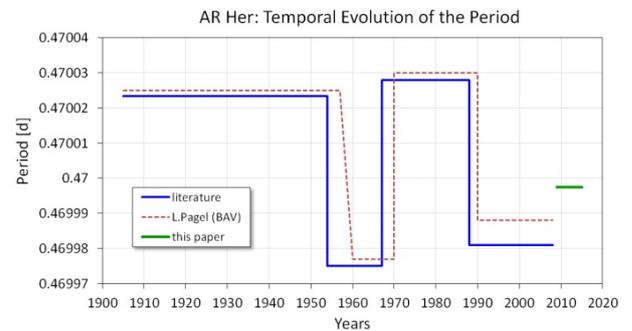

**Figure 1:**   Temporal evolution of the period according to the values from the literature, which have been complemented by the results from Pagel [2008] and the present investigation (see text for details).

**Blazhko Effect |** Balázs and Detre [1939] investigated the variations in maximum and minimum brightness of AR Her. During the period 1935 to 1939, the maximum photographic brightness was found to vary by an amount of 0.53 mag (10.28–10.81 mag) and the minimum photographic brightness by an amount of 0.34 mag (11.71–12.05 mag), resulting in amplitude variations between 0.90 mag and 1.77 mag. Balázs and Detre [1939] estimated a Blazhko period of $P_B$ = 31.49 days.

Almar [1961] reached the conclusion that the observed changes in period are very irregular and hence difficult to interpret. He determined a slightly longer Blazhko period of $P_B$ = 31.5489 days.

An even larger Blazhko period value was given by Wils et al. [2006], who derived $P_B$ = 32 days.

## 2 Observations

**Data |** 34,102 brightness measurements, which were won by 31 observers in the time span from HJD 2438506 to HJD 2457277, were extracted from the AAVSO International Database. Only V and TG (tricolor green) measurements were included into the analysis. The data were searched for time intervals suited to the timing of maximum light and the determination of the corresponding brightness.

**Timings of Maximum Light |** Sufficiently short time intervals were chosen, which allowed the determination of times of maximum brightness by a polynomial regression from $2^{nd}$ to $6^{th}$ order ($3^{rd}$ order was applied in 70% of cases).

The resulting graphs were inspected visually; in two cases, the computed times of maximum brightness had to be adjusted slightly (the corresponding datasets were charaterized by very sharp maxima). Whenever possible, the derived times of maximum brightness were compared to the corresponding values determined by the AAVSO observers. A very good agreement between the two sources was found, which confirms the values derived in the present investigation.

The derived times of maximum brightness are listed in Appendix 1.

**Investigated Time Span |** All derived maxima stem from the time interval HJD 2452745 to 2457245 (14. 04. 2003 – 10. 08. 2015). Prior to HJD 2454880 (17. 02. 2009), the maxima follow the (O–C) trend that has been shown to exist since about 1990 [Pagel 2008]. The present work investigates the behaviour of the period from 2009 to 2015; during this time span, a longer period was detected.

103 times of maximum were derived from AAVSO data collected by the following 14 observers:

| | |
|---|---|
| Arranz Lázaro, Alejandra (AALB) | 1 |
| Arranz, Teofilo (ATE) | 27 |
| Bialozynski, Jerry (BIZ) | 1 |
| Boardman, James (BJAA) | 1 |
| Banfi, Massimo (BVN) | 4 |
| Dvorak, Shawn (DKS) | 8 |
| Hambsch, Franz-Josef (HMB) | 2 |
| Menzies, Kenneth (MZK) | 3 |
| Papini, Riccardo (PCC) | 5 |
| Pagel, Lienhard (PLN) | 18 |
| Poklar, Rudy (PRX) | 2 |
| Petriew, Vance (PVA) | 1 |
| Samolyk, Gerard (SAH) | 22 |
| Sabo, Richard (SRIC) | 8 |

Another 13 maxima were added by L. Pagel, which had been gleaned from *BAV Mitteilungen* [Hübscher, 2016b] and were not incorporated into the AAVSO database.

## 3 Results and Discussion

### 3.1 Revised Elements

Using the elements provided by the GCVS and the VSX (Equation 1), the resulting (O–C) diagram indicates a jump in epoch, which is caused by the period being slightly too long. Thus, the resulting discrepancies in (O–C) add up to a full cycle during the investigated time span.

In order to avoid this, the elements from Hübscher [2016a] were employed (Equation 2). The resulting diagram is shown in Figure 2.

From the weighted best-fit line

(3) O–C = –0.0115$^d$ + 9.90·10$^{-6}$ ·E

in Figure 2, the following improved elements have been derived:

(4)  C = HJD 2454683.4251 +0.46999746 · E
             ±0.0058  ±0.00000145

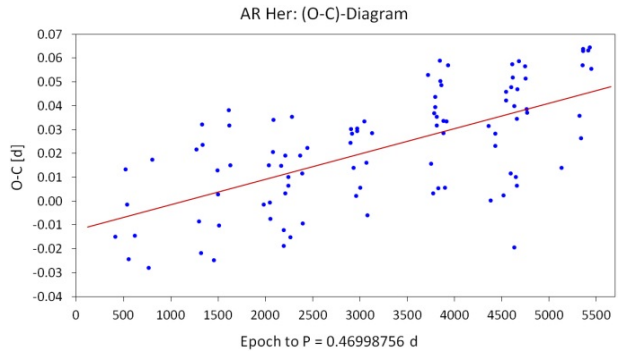

**Figure 2:** (O–C) values as a function of epoch according to Equation (2).

The indicated error has been based on a weighted linear regression. It is noteworthy to point out that the variations due to the Blazhko effect influence the error values. On the basis of the calculations of the Blazhko period value (see below), the actual period uncertainty is estimated to be approximately half the above listed value.

### 3.2 Variations in the Time of Maximum

Figure 3 illustrates the changes in (O–C) as a function of Julian Date according to Equation (4), which are used for the analysis of the Blazhko effect. It has to be pointed out that part of the observed scattering is also due to measurement errors.

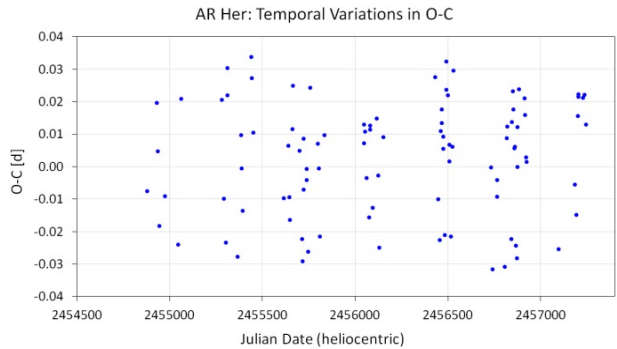

**Figure 3:** (O–C) values as a function of HJD according to Equation (4). The covered time span is February 2009 to August 2015.

The (O–C) values are scattered between −0.0317 and 0.0336 days, which corresponds to variations in maximum time of −6.74 % to 7.15 % of the period due to the Blazhko effect.

### 3.3 Variations in Maximum Brightness

Figure 4 illustrates the temporal evolution of the visual maximum brightness $V_{max}$.

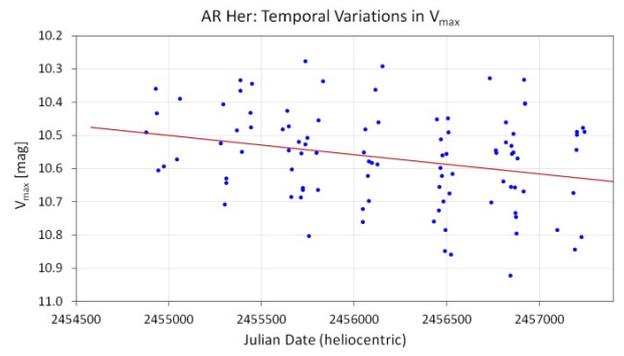

**Figure 4:** Temporal evolution of the visual maximum brightness $V_{max}$ in the timespan February 2009 to August 2015.

In the time span 2009 to 2015, the visual maximum brightness varied between 10.28 mag < $V_{max}$ < 10.92 mag, resulting in a mean visual maximum brightness of $V_{max}$ = 10.569 mag.

Assuming $m_{pg} \approx B$ and a mean (B–V) index of 0.17 mag [Weigert, 2005], the photographic maximum brightness varied between 10.1 mag < $V_{max}$ < 10.6 mag during the time span 1935 to 1939 [Balázs, 1939].

The more recently determined values are considerably fainter. In comparison to the mean maximum brightness of 10.59 mag given by Kholopov [1988], the corresponding value determined in the present work is slightly brighter.

The long-term changes in maximum brightness are readily discernible during the time span 2009 to 2015, in which the maximum brightness continually decreased by 22 mmag/year.

### 3.4 Blazhko Effect of the Time of Maximum

In order to determine the Blazhko period $P_B$, the value set from Fig. 3 has been subjected to a Fourier analysis employing the program package Period04 [Lenz, 2005]. The best fit results in a Blazhko period of $P_B$ = 32.062 ±0.013 day and the following equation:

(5) $\quad$ O–C = −0.0017(±16) + 0.0241(±13) · sin (2π · (0.03118961(±122) · t + 0.923(±8)))

where the time (t) is given in HJD and the errors of the coefficients, which apply to the last digits, are listed in parantheses.

From Equations (4) and (5), the following, improved elements have been derived, which take into account the Blazhko period.

(6)  C =  HJD 2457245.3829 + 0.46999746 · E
         + 0.0241 · sin(0.092105456·E + 2.8519)

Figure 5 illustrates the residual errors in (O–C). A further analysis indicates that, with the available data, no significant improvement is possible.

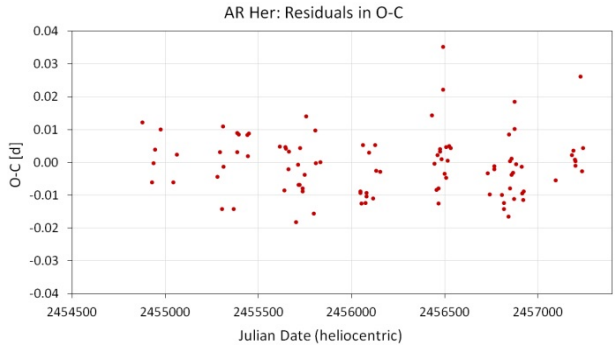

**Figure 5:**  Residuals in (O–C) according to Equation (5).

The squared deviation of the value sets of Figures 3 and 5 mirrors the quality and relevance of the Blazhko period derived from the (O–C) values:

   not taking into account $P_B$:   0.0174 days (Fig. 3)
   taking into account $P_B$:       0.0093 days (Fig. 5)

**Phase Diagram |** Figure 6 shows the phase diagram folded on the Blazhko period $P_B$ = 32.062 days. It visualizes the character of the Blazhko effect whose asymmetric nature is only approximately described by the best fit sine curve.

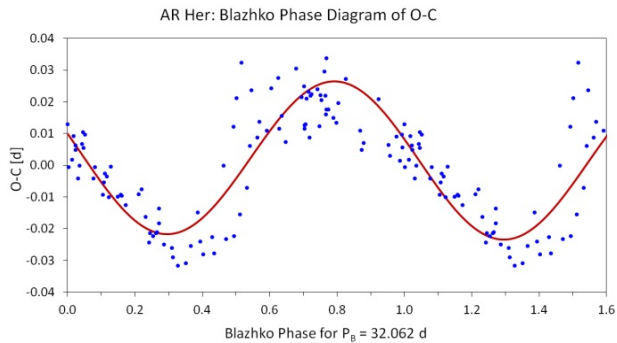

**Figure 6:**  Blazhko phase diagram of the (O–C) values, folded with a Blazhko period of $P_B$ = 32.062 days.

**Period Control |** In order to confirm the derived Blazhko period, a self-developed period search program was employed, which calculates the period from the least squares deviation of a polynomial fit to the data. A Blazhko period of $P_B$ = 32.054 ±0.001 days was calculated from the (O–C) values, which is in agreement with the above listed period value.

## 3.5 Blazhko Effect of the Maximum Brightness

The Blazhko period was derived from the visual maximum brightness $V_{max}$ using the program package Period04. To this end, the differences in maximum brightness $\Delta V_{max}$ (Figure 7) were calculated from the differences between the measured maximum brightness and the best-fit line in Figure 4.

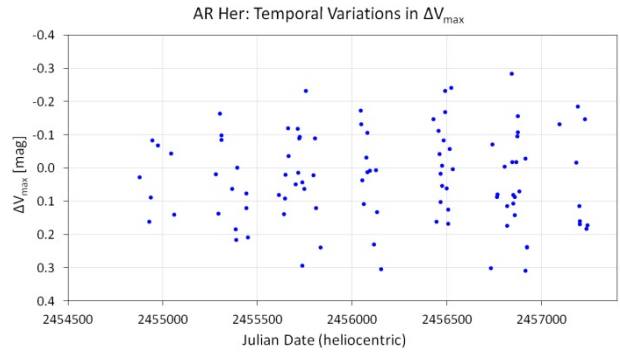

**Figure 7:**  Temporal evolution of the differences in maximum brightness $\Delta V_{max}$ between the measured maximum brightness $V_{max}$ and the best-fit line in Figure 4.

The best fit to the values in Figure 7 results in a Blazhko period of $P_B$ = 29.528 ±0.017 days. The residuals are shown in Figure 8.

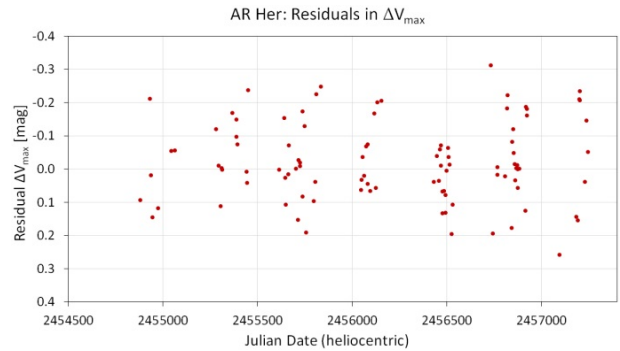

**Figure 8:**  Residuals in $V_{max}$ after fitting the data with a Blazhko period of $P_B$ = 29.528 ±0.017 days.

A further analysis of the residual values resulted in no significant improvement.

The square deviation of the value sets of Figures 7 and 8 mirrors the quality and relevance of the Blazhko period derived from $V_{max}$:

   not taking into account $P_B$:   0.135 mag (Fig. 7)
   taking into account $P_B$:       0.120 mag (Fig. 8)

**Phase Diagram |** Figure 9 shows the phase diagram folded on the Blazhko period $P_B$ = 29.528 days and visualizes the character of the Blazhko effect. The

resultant curve is symmetric, which is in contrast to the Blazhko phase diagram based on the (O–C) values (Figure 6). A further analysis indicates that, with the available data, no significant improvement is possible.

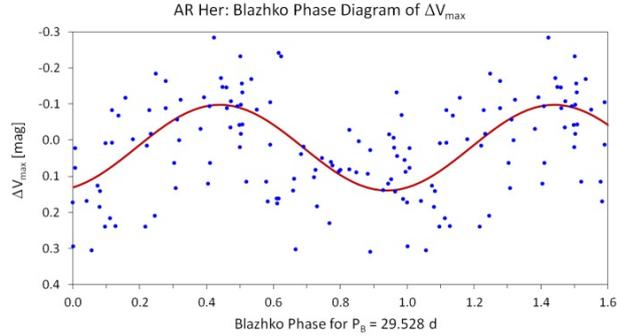

**Figure 9:** Blazhko phase diagram of $V_{max}$, folded with a Blazhko period of $P_B$ = 29.528 days.

**Period Control** With the help of a self-developed period search program, which calculates the period from the least squares deviation of a polynomial fit to the data, a Blazhko period of $P_B$ = 29.53 ±0.01 days was calculated from $V_{max}$, which is in agreement with the above listed period value.

### 3.6 (O–C) vs. $V_{max}$ Diagram

Poretti et al. [2016] have shown that, in RR Lyrae stars, there often is a correlation between the (O–C) values and $V_{max}$. When plotting $V_{max}$ versus (O–C) in a diagram, a closed curve resembling an ellipse is derived, often called 'Blazhko potatoe'. Certain $V_{max}$ values are observed at different (O–C) values and vice versa, and the resultant curve can be run in opposite directions, i.e. clockwise or anticlockwise [Le Borgne, 2012].

**Simulation |** Several simulations were run in order to get a better understanding of the processes in the Blazhko (O–C) vs. $V_{max}$ diagram. If the Blazhko periods derived from the (O–C) values and $V_{max}$ are identical, the resultant curve is a closed ellipse (Figure 10, left panel). The orientation of the line of apsides depends on the phases of both Blazhko periods.

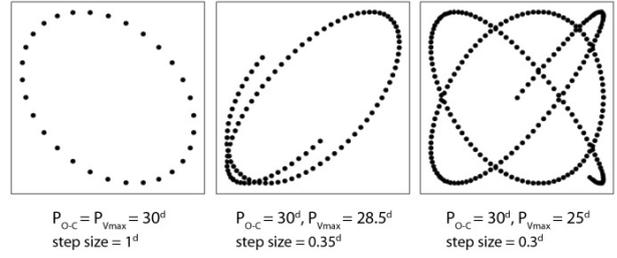

**Figure 10:** Simulation of Blazhko (O–C) vs. $V_{max}$ diagrams assuming different Blazhko periods. The step size only has implications on the density of points.

If both Blazhko periods are different, the resultant ellipses are not closed but open (Figure 10, middle panel). Furthermore, the line of apsides is rotating. This becomes obvious in the case of large differences between both Blazhko periods (Figure 10, right panel): The curve resembles an open ellipse and the line of apsides is rotating strongly. In addition to that, changes in the eccentricity of the ellipse are apparent.

The above mentioned periodic changes are here described with a period I will refer to as 'Blazhko super period' $P_{Super}$:

$$(7) \quad P_{Super} = \frac{P_{O-C} \cdot P_{V_{max}}}{|P_{O-C} - P_{V_{max}}|}$$

In the course of a Blazhko super period, the direction of rotation changes twice. This change takes place when the ellipse is transformed to a straight line. The work of Poretti et al. demonstrates that the resultant curves can be run in opposite directions, i.e. clockwise or anticlockwise.

The examples shown in Figure 10 are symbolic simulations. Nevertheless, they have been based on values which are close to what is actually observed in RR Lyrae stars.

Figure 11 illustrates several Blazhko (O–C) vs. $V_{max}$ diagrams that have been calculated from the periods of AR Herculis as determined in the present work ($P_{O-C}$ = 32.062 days, $P_{V_{max}}$ = 29.528 days). Additional scatter was added which corresponds to (and thus mimics) the mean measured errors in (O–C) (±0.0039 days ≈ 13 %) and $V_{max}$ (±0.0047 mag ≈ 3%).

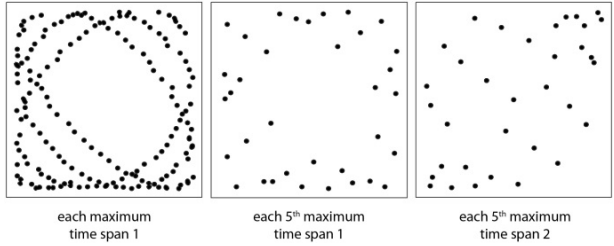

**Figure 11:** Simulation of Blazhko (O–C) vs. $V_{max}$ diagrams for AR Herculis, based on the periods of 32.062 days and 29.528 days. Additional scatter was added which corresponds to the mean measured errors in (O–C) (±0.0039 days) and $V_{max}$ (±0.0047 mag). The left panel illustrates each maximum during a time span of 100 days. The middle panel has been based on each fifth maximum in the same time span. The right panel shows each fifth maximum during a different 100-day interval.

Because of the rotation of the line of apsides and the changes in eccentricity, an ellipsoidal curve is present only in a short time interval like e.g. three Blazhko periods. In the case of AR Herculis, this corresponds to about 100 days. The first simulated plot (Figure 11, left panel) illustrates the positions of the maxima, taking into account each single maximum. The second simulation (Figure 11, middle panel) has been based on each fifth maximum only (assuming an even distribution). From the resulting loose cloud of points, the ellipsoidal shape can hardly be divined. The third simulation (Figure 11, right panel) is equivalent to the second one but has been based on a different 100-day time interval.

The presence of a discernible ellipse in the Blazhko (O–C) vs. $V_{max}$ diagram of an RR Lyrae star (as e.g. AR Herculis) is therefore dependent on the difference of the observed periods, the density of maximum timings, and the time interval of the Blazhko super period.

In the case of AR Herculis, the observed Blazhko periods of 32.062 days and 29.528 days result in a Blazhko super period of 373.6 days. Thus, the observed direction of rotation should change every 186.8 days.

In order to investigate this, the maximum timings of AR Her have been plotted in Figure 12. After plotting all observed maxima collected over a time span of 2,366 days, no ellipsoidal shape is discernible. To tackle the question if time intervals of limited duration will produce ellipsoidal curves, the observations have been divided into 7 time intervals, which were analysed separately. The time intervals are defined by the observational gaps in the data and are obvious in Figures 3 and 7.

Indications of correlated behaviour are present in the time interval around HJD 2456500 (blue dots in Figures 12 and 13).

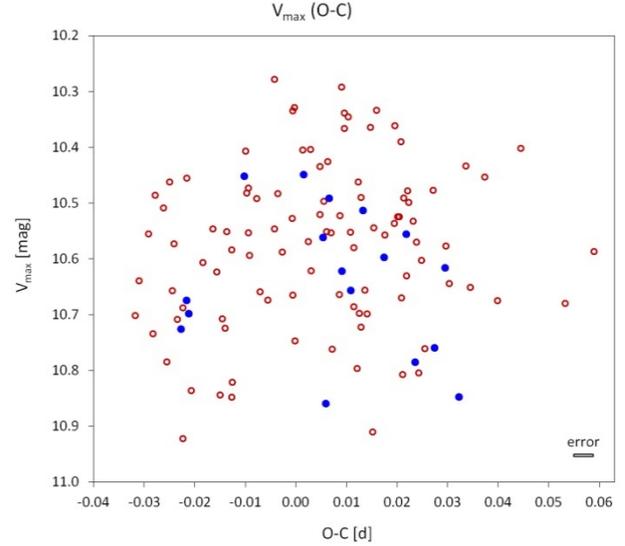

**Figure 12:** Diagram for AR Her, which plots $V_{max}$ vs. (O–C). The blue dots represent a time interval of limited duration. The mean error of the observations is indicated by the rectangle in the bottom right corner.

After arranging the measurements chronologically and connecting them with a line, there are indications of a counterclockwise rotation.

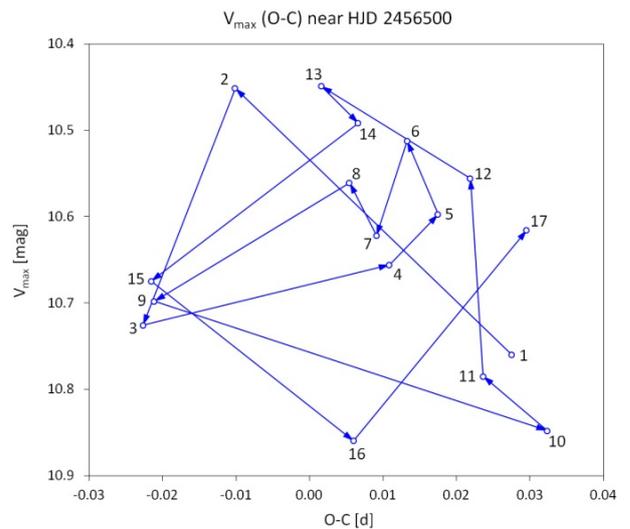

**Figure 13:** Diagram for AR Her, which plots $V_{max}$ vs. (O–C) for the time span HJD 2456432 to 2456532. The chronological order of the measurements is indicated.

Assuming the presence of a datapoint between measurements #3 and #4 (at HJD 2456460±1, near measurement #16 in the plot), a roughly circular shape is conceivable.

The represented period of 100 days encompasses three complete cycles, i.e. one cycle is about 32 days – a value, which corresponds exactly to the determined Blazhko period.

## 4   Conclusion

For the time span 2009 to 2015, a new period value of $P = 0.46999746(\pm145)$ d was derived from the (O–C) analysis. This confirms the irregular variations in period that have been observed for more than 100 years in AR Herculis. Apparently, the period is apt to change in the time span of a few months, after which it remains practically constant for several years.

An analysis of the derived maximum timings indicates a strong Blazhko effect with a Blazhko period of $P_B = 32.062(\pm13)$ days. The evolution of the observed maximum brightness $V_{max}$, which decreased from 2009 to 2015 by 22 mmag/year, also indicates the presence of a Blazhko effect with $P_B = 29.528(\pm17)$ days.

Thus, the Blazhko period derived from $V_{max}$ is about 8% shorter; it is also subject to long-term variations (declining at present). Therefore, 11.65 (O–C) Blazhko periods fit into 12.65 $V_{max}$ Blazhko periods, which results in a Blazhko super period of 373.6 days.

This value corresponds to the period in the Blazhko (O–C) vs. $V_{max}$ diagram, which manifests itself e.g. through the observed change of the direction of rotation halfway through the period. With the same period, the line of apsides rotates and the eccentricity of the (open) ellipse changes. The change of the direction of rotation takes place when the ellipse is transformed to a straight line.

I maintain a critical view regarding the differences between the Blazhko periods derived from (O–C) and $V_{max}$. A physical explanation of this phenomenon is lacking, which also holds true for the Blazhko effect in general.

As is the case for other RR Lyrae stars, the positions of the maxima in the (O–C) vs. $V_{max}$ diagram for AR Herculis should describe an ellipsoidal curve, with the direction of rotation changing after about half a year. However, a conceivable ellipse is present only in a single time segment of 100 days. Clearly, further studies are necessary in this respect.

In order to shed more light on this phenomenon, as many maxima as possible (*N*>50) need to be observed during a three-month period. In doing so, it is important to exactly determine date and brightness of the observations.

From the determination of at least 100 maxima during the course of a year, a statement could be made on the eccentricity, the rotation of the line of apsides, and the direction of rotation.

## Acknowledgements

The author would like to thank the American Association of Variable Star Observers (AAVSO) for maintaining the AAVSO International Database and the above listed observers who have provided carefully taken measurements which have been employed in the present work. Furthermore, observations from the Bundesdeutsche Arbeitsgemeinschaft für Veränderliche Sterne (BAV) were employed. Peter B. Lehmann pointed me to the work of Poretti et al., which initiated the present research. Thanks are due to Prof. Dr. Lienhard Pagel for his critical comments which helped to improve the paper. Stefan Hümmerich is acknowledged for reviewing and correcting the manuscript.

## References

Almar, I.: MiBud **51** (1961)

Balázs, Julia & L.Detre: MiBud **8** (1939)

Feast, M.W. et al.: MNRAS **386** (2008), 2115–2134

Hübscher, J. [a]: BAV Circular, Heft 1-2016

Hübscher, J. [b]: Database Rev. 13, BAV Mitteilungen (accessed 01/02/2016)

Kholopov, P.N. et al.: GCVS4, I–III (1985–1988)

Klepikova, L.A.: PZ**11**, No. 3 (1957), 137

Lange, G.A.: ATsir **503** (1969), 7–8

Le Borgne, J.-F. et al.: AJ **144** (2012), 39

Lenz, P. & M. Breger: CoAst **146** (2005), 53

Pagel, L.: BAV Rundbrief, Heft 3-2008

Poretti, E. et al.: arXiv:1601.02772v1 (2016)

The AAVSO International Database (AID), www.aavso.org/data-download (accessed 01/02/2016)

The International Variable Star Index (VSX) of the AAVSO, www.aavso.org/vsx (accessed 06/02/2016)

Weigert, A. et al.: ISBN 3-527-40358-2 (2005)

Wils, P. et al.: MNRAS **368** (2006), 1757–1763

# Appendix

Time [HJD] and brightness [mag] of the calculated times of maximum in the time span 2003–2015. (O–C) values are calculated according to R = HJD 2454683.4251 + 0.46999746·E.

| HJD | Error | $V_{max}$ | Error | O–C | HJD | Error | $V_{max}$ | Error | (O–C) |
|---|---|---|---|---|---|---|---|---|---|
| 2452744.7152 | 0.0032 | 10.5766 | 0.0024 | 0.0297 | 2456054.8884 | 0.0035 | 10.5525 | 0.0027 | 0.0108 |
| 2453489.6905 | 0.0038 | 10.5863 | 0.0018 | 0.0590 | 2456062.8640 | 0.0040 | 10.4830 | 0.0026 | -0.0036 |
| 2453497.6659 | 0.0018 | 10.4013 | 0.0008 | 0.0444 | 2456075.5419 | 0.0139 | 10.6230 | 0.0030 | -0.0156 |
| 2453505.6309 | 0.0035 | 10.5361 | 0.0023 | 0.0195 | 2456081.6789 | 0.0050 | 10.5795 | 0.0080 | 0.0114 |
| 2453510.7844 | 0.0021 | 10.6209 | 0.0013 | 0.0030 | 2456081.6800 | 0.0044 | 10.6975 | 0.0040 | 0.0125 |
| 2453519.7646 | 0.0056 | 10.6795 | 0.0034 | 0.0533 | 2456096.6947 | 0.0029 | 10.5840 | 0.0020 | -0.0127 |
| 2453536.6514 | 0.0006 | 10.5244 | 0.0001 | 0.0201 | 2456116.4621 | 0.0035 | 10.3640 | 0.0030 | 0.0148 |
| 2453817.7271 | 0.0038 | 10.4528 | 0.0053 | 0.0374 | 2456126.7845 | 0.0056 | 10.5878 | 0.0057 | -0.0028 |
| 2454245.3749 | 0.0034 | 10.8209 | 0.0031 | -0.0125 | 2456132.4022 | 0.0014 | 10.4620 | 0.0030 | -0.0250 |
| 2454255.7672 | 0.0129 | 10.6751 | 0.0162 | 0.0398 | 2456154.5261 | 0.0035 | 10.2920 | 0.0030 | 0.0090 |
| 2454260.4618 | 0.0022 | 10.6510 | 0.0012 | 0.0345 | 2456431.8431 | 0.0065 | 10.7601 | 0.0028 | 0.0275 |
| 2454276.3946 | 0.0023 | 10.8476 | 0.0014 | -0.0127 | 2456447.7854 | 0.0021 | 10.4519 | 0.0020 | -0.0101 |
| 2454300.3912 | 0.0022 | 10.6984 | 0.0012 | 0.0141 | 2456457.6428 | 0.0034 | 10.7262 | 0.0013 | -0.0227 |
| 2454596.4548 | 0.0034 | 10.8359 | 0.0033 | -0.0207 | 2456462.8463 | 0.0019 | 10.6565 | 0.0008 | 0.0109 |
| 2454636.4405 | 0.0031 | 10.9100 | 0.0018 | 0.0152 | 2456468.4929 | 0.0042 | 10.5978 | 0.0042 | 0.0175 |
| 2454642.5608 | 0.0038 | 10.7611 | 0.0022 | 0.0255 | 2456469.4287 | 0.0035 | 10.5131 | 0.0050 | 0.0133 |
| 2454658.5006 | 0.0025 | 10.7073 | 0.0014 | -0.0146 | 2456476.4745 | 0.0090 | 10.6222 | 0.0015 | 0.0091 |
| 2454660.3812 | 0.0031 | 10.7239 | 0.0015 | -0.0140 | 2456477.4108 | 0.0032 | 10.5615 | 0.0027 | 0.0054 |
| 2454879.8763 | 0.0117 | 10.4919 | 0.0259 | -0.0077 | 2456484.4342 | 0.0025 | 10.6989 | 0.0016 | -0.0211 |
| 2454930.6633 | 0.0040 | 10.3608 | 0.0043 | 0.0196 | 2456492.4776 | 0.0045 | 10.8483 | 0.0021 | 0.0323 |
| 2454937.6984 | 0.0024 | 10.4340 | 0.0024 | 0.0047 | 2456493.4089 | 0.0019 | 10.7854 | 0.0005 | 0.0236 |
| 2454945.6653 | 0.0031 | 10.6062 | 0.0021 | -0.0183 | 2456500.4571 | 0.0022 | 10.5564 | 0.0011 | 0.0219 |
| 2454975.7543 | 0.0046 | 10.5938 | 0.0042 | -0.0092 | 2456508.4268 | 0.0034 | 10.4492 | 0.0068 | 0.0016 |
| 2455045.7690 | 0.0045 | 10.5732 | 0.0043 | -0.0241 | 2456509.3718 | 0.0015 | 10.4920 | 0.0012 | 0.0066 |
| 2455062.7338 | 0.0026 | 10.3899 | 0.0025 | 0.0208 | 2456516.3936 | 0.0012 | 10.6751 | 0.0003 | -0.0216 |
| 2455281.7523 | 0.0046 | 10.5244 | 0.0051 | 0.0205 | 2456525.3511 | 0.0045 | 10.8600 | 0.0069 | 0.0060 |
| 2455294.4118 | 0.0015 | 10.4071 | 0.0010 | -0.0100 | 2456532.4246 | 0.0020 | 10.6160 | 0.0010 | 0.0295 |
| 2455304.7383 | 0.0033 | 10.7084 | 0.0015 | -0.0234 | 2456733.5537 | 0.0035 | 10.3290 | 0.0030 | -0.0003 |
| 2455311.3720 | 0.0025 | 10.6438 | 0.0005 | 0.0303 | 2456742.9222 | 0.0071 | 10.7019 | 0.0110 | -0.0317 |
| 2455312.7735 | 0.0043 | 10.6299 | 0.0052 | 0.0218 | 2456766.9196 | 0.0017 | 10.5464 | 0.0016 | -0.0042 |
| 2455367.7135 | 0.0046 | 10.4855 | 0.0041 | -0.0279 | 2456767.8544 | 0.0067 | 10.5528 | 0.0181 | -0.0094 |
| 2455387.4908 | 0.0047 | 10.3663 | 0.0076 | 0.0095 | 2456807.7826 | 0.0023 | 10.6388 | 0.0014 | -0.0310 |
| 2455388.4206 | 0.0027 | 10.3341 | 0.0034 | -0.0006 | 2456819.5722 | 0.0035 | 10.5220 | 0.0030 | 0.0087 |
| 2455394.5175 | 0.0053 | 10.5508 | 0.0058 | -0.0137 | 2456820.5158 | 0.0035 | 10.4620 | 0.0030 | 0.0123 |
| 2455442.5046 | 0.0015 | 10.4333 | 0.0005 | 0.0336 | 2456844.4510 | 0.0010 | 10.9220 | 0.0010 | -0.0224 |
| 2455444.3781 | 0.0030 | 10.4764 | 0.0007 | 0.0272 | 2456846.8370 | 0.0037 | 10.6562 | 0.0019 | 0.0136 |
| 2455451.4113 | 0.0068 | 10.3452 | 0.0182 | 0.0104 | 2456851.5465 | 0.0017 | 10.5318 | 0.0008 | 0.0232 |
| 2455615.4203 | 0.0020 | 10.4823 | 0.0011 | -0.0097 | 2456853.4209 | 0.0061 | 10.5566 | 0.0146 | 0.0176 |
| 2455640.8162 | 0.0110 | 10.4259 | 0.0463 | 0.0063 | 2456860.4588 | 0.0037 | 10.4966 | 0.0056 | 0.0055 |
| 2455647.3804 | 0.0027 | 10.4733 | 0.0026 | -0.0095 | 2456861.3994 | 0.0019 | 10.5513 | 0.0017 | 0.0061 |
| 2455649.7234 | 0.0064 | 10.5459 | 0.0123 | -0.0164 | 2456868.4188 | 0.0010 | 10.6570 | 0.0010 | -0.0245 |
| 2455662.4412 | 0.0042 | 10.6855 | 0.0026 | 0.0114 | 2456873.5850 | 0.0035 | 10.7340 | 0.0030 | -0.0282 |
| 2455664.8046 | 0.0041 | 10.6024 | 0.0220 | 0.0248 | 2456875.4930 | 0.0035 | 10.7470 | 0.0030 | -0.0002 |
| 2455702.3843 | 0.0020 | 10.5200 | 0.0005 | 0.0047 | 2456876.4453 | 0.0010 | 10.7961 | 0.0020 | 0.0121 |
| 2455714.5771 | 0.0033 | 10.6875 | 0.0031 | -0.0224 | 2456884.4470 | 0.0010 | 10.5703 | 0.0020 | 0.0238 |
| 2455716.4503 | 0.0024 | 10.5549 | 0.0013 | -0.0292 | 2456915.4639 | 0.0035 | 10.6700 | 0.0030 | 0.0209 |
| 2455723.5223 | 0.0067 | 10.6590 | 0.0039 | -0.0071 | 2456917.3389 | 0.0035 | 10.3330 | 0.0030 | 0.0159 |
| 2455724.4780 | 0.0030 | 10.6643 | 0.0010 | 0.0086 | 2456923.4358 | 0.0035 | 10.4040 | 0.0030 | 0.0028 |
| 2455738.5686 | 0.0085 | 10.5275 | 0.0121 | -0.0008 | 2456924.3743 | 0.0035 | 10.4050 | 0.0030 | 0.0014 |
| 2455739.5051 | 0.0051 | 10.2777 | 0.0121 | -0.0043 | 2457096.8365 | 0.0010 | 10.7850 | 0.0020 | -0.0255 |
| 2455748.4131 | 0.0020 | 10.5087 | 0.0012 | -0.0262 | 2457184.7460 | 0.0060 | 10.6742 | 0.0067 | -0.0055 |
| 2455757.8635 | 0.0050 | 10.8040 | 0.0040 | 0.0242 | 2457193.6665 | 0.0110 | 10.8437 | 0.0096 | -0.0150 |
| 2455798.7360 | 0.0024 | 10.5530 | 0.0020 | 0.0070 | 2457201.6869 | 0.0046 | 10.5439 | 0.0121 | 0.0154 |
| 2455806.7184 | 0.0020 | 10.6645 | 0.0009 | -0.0006 | 2457203.5728 | 0.0060 | 10.4905 | 0.0096 | 0.0214 |
| 2455810.4574 | 0.0023 | 10.4550 | 0.0020 | -0.0216 | 2457204.5137 | 0.0029 | 10.4985 | 0.0022 | 0.0223 |
| 2455834.4584 | 0.0026 | 10.3382 | 0.0025 | 0.0096 | 2457229.4224 | 0.0038 | 10.8070 | 0.0018 | 0.0211 |
| 2456047.8349 | 0.0057 | 10.7618 | 0.0071 | 0.0072 | 2457237.4133 | 0.0026 | 10.4783 | 0.0023 | 0.0220 |
| 2456049.7205 | 0.0036 | 10.7218 | 0.0032 | 0.0128 | 2457245.3941 | 0.0008 | 10.4897 | 0.0005 | 0.0129 |